\documentclass[usenatbib]{mn2e}
\usepackage{graphicx}

\title[A Library of Lick/IDS Indices for BSPs]
{A Library of Lick/IDS Indices for Binary Stellar Populations}

\author[F. Zhang and L. Li]
{Fenghui~Zhang\thanks{E-mail: gssephd@public.km.yn.cn;
zhang\_fh@hotmail.com} and Lifang~Li\\
National Astronomical Observatories/Yunnan Observatory, Chinese
Academy of Sciences, PO Box 110, Kunming, Yunnan Province,\\
650011, China \\}

\begin{document}

\date{\today}

\pagerange{\pageref{firstpage}--\pageref{lastpage}}

\pubyear{2005}

\maketitle

\label{firstpage}

\begin{abstract}
Using evolutionary population synthesis, we present 13 refined
absorption-line indices defined by the Lick Observatory Image
Dissector Scanner (Lick/IDS) system for an extensive set of
instantaneous-burst binary stellar populations (BSPs) at high
resolution ($\sim 0.3\, \rm \AA$), and 38 indices at intermediate
resolution ($3\, \rm \AA$). The ages of the populations are at an
interval of 1\,Gyr in the range 1--15\,Gyr, and the metallicities
are in the range 0.004--0.03. These indices are obtained by two
methods: (1) obtain them by using the empirical fitting functions
(FFs method); (2) measure them directly from the synthetic spectra
(DC method). Together with our previous paper a database of
Lick/IDS spectral absorption-line indices for BSPs at high and
intermediate resolutions is provided. This set of indices includes
21 indices of \citet{wor94}, four Balmer indices defined by
\citet{wor97} and 13 indices with the new passband definitions of
\citet{tra98}. The full set of synthetic indices and the
integrated pseudo-continuum are listed in the Appendix, which is
only available online at http://www.blackwellpublishing.com/... or
from our website (http://www.ast9.com/), or on request from the
first author. Moreover, the full set of the integrated spectral
energy distributions can be obtained from our website.

We compare the synthetic Lick/IDS indices obtained by FFs method
and those by DC method for BSPs with various metallicities, and
find that the discrepancies are significant: Ca4455 (index 6),
Fe4668 (8), Mg$_{\rm b}$ (13), Fe5709 (17), NaD (19), TiO$_1$ (20)
and TiO$_2$ (21, except for $Z=0.02$) in the W94 system,
Ca4455$\rm^T$ (6$\rm^T$), C$_2$4668$\rm^T$ (8$\rm^T$), NaD$\rm^T$
(19$\rm^T$), TiO$_1\rm^T$ (20$\rm^T$) and TiO$_2\rm^T$ (21$\rm^T$,
except for $Z=0.02$) in the T98 system obtained by DC method are
less (bluer) than the corresponding ones obtained by FFs method
for all metallicities. Ca4227 (index 3), Fe5782 (18),
Ca4227$\rm^T$ (3$\rm^T$) and Fe5782$\rm^T$ (18$\rm^T$) are greater
at $Z=0.03$ and become to be bluer at $Z=0.004$, Fe5709 $\rm^T$
(17$\rm^T$) index is less at $Z=0.03$ and becomes to be redder at
$Z=0.004$ than the corresponding ones obtained by FFs method.
\end{abstract}


%1__________________________________________________________________
\section{Introduction}
Using evolutionary stellar synthesis (EPS), we presented 21
absorption-line indices (following the definitions of
\citealt{wor94}, hereafter W94) defined by the Lick Observatory
Image Dissector Scanner (Lick/IDS) system at low spectral
resolution \citep[10--20\,$\rm \AA$,][hereafter Paper I]{zha05a},
and 25 Lick/IDS absorption-line indices (21 indices of W94 and
four Balmer indices defined by \citealt{wor97}, hereafter WO97) at
high resolution \citep*[$\sim 0.3\,\rm \AA$,][hereafter Paper
II]{zha05b}, for an extensive set of instantaneous-burst binary
stellar populations (BSPs) with binary interactions.

In Lick/IDS absorption-line index system, \citet[][hereafter
T98]{tra98} refined the wavelength definitions of the 13 indices
in the W94 system. As an UPDATE of Paper II, this paper will
present the 13 Lick/IDS indices refined by T98 for BSPs at high
resolution ($\sim 0.3\, \rm \AA$). To complete this library we
also present these indices for BSPs at intermediate resolution
($3\, \rm \AA$). A full model description and algorithm are given
in Papers I and II.

The outline of the paper is as follows: the results of the 13
refined Lick/IDS absorption-line indices for BSPs at high
resolution in Section 2, the results at intermediate resolution in
Section 3, in Section 4, we give summary and conclusions.

%2__________________________________________________________________
\section{Lick/IDS indices for BSPs at high resolution ($\sim 0.3\, \rm \AA$)}
\begin{figure*}
\centering{
\includegraphics[bb=77 100 571 690,height=17.1cm,width=17.5cm,clip,angle=-90]{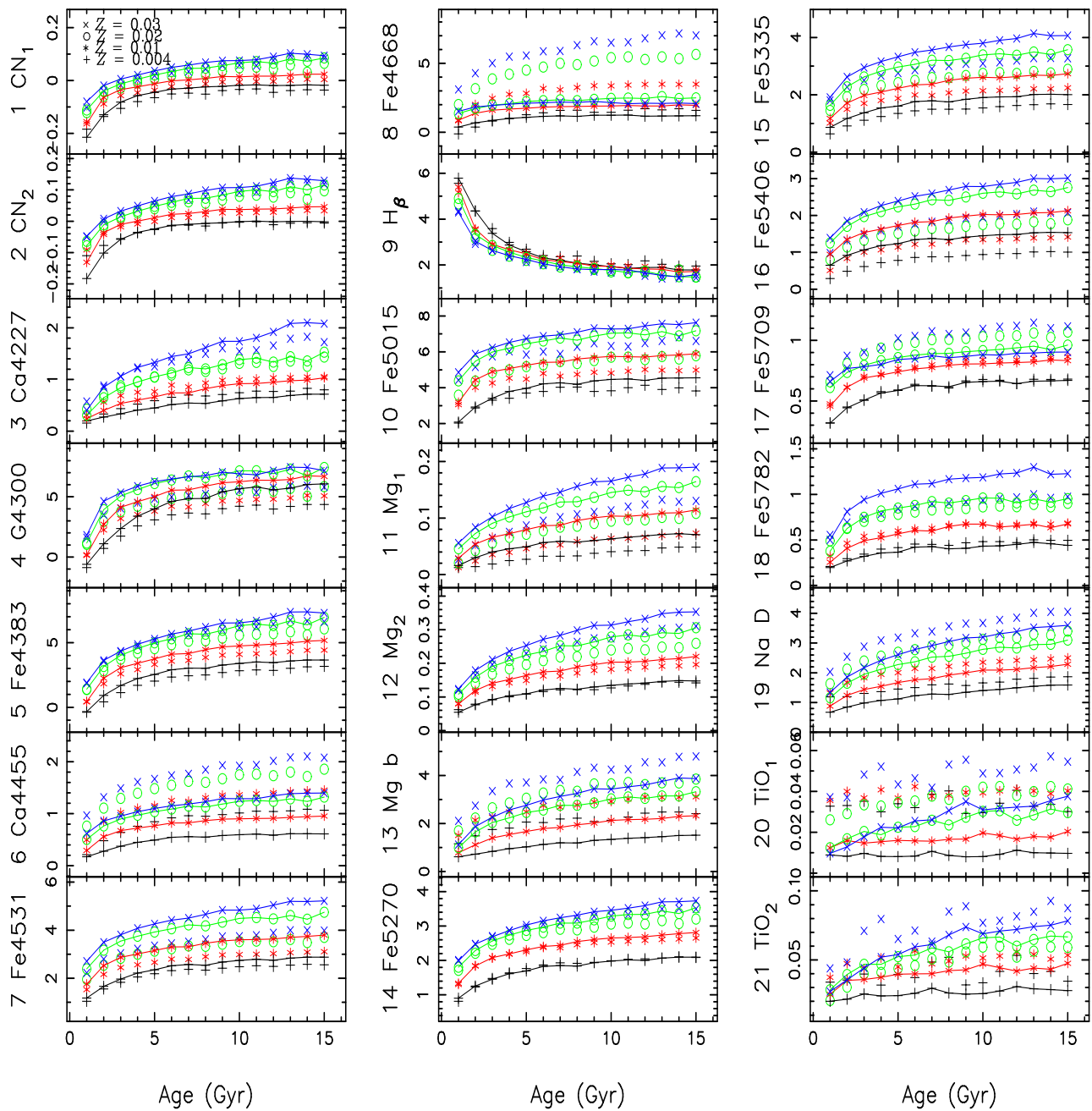}
} \caption{Lick/IDS absorption-line indices in the W94 system
obtained via DC (line+symbols) and FFs (symbols) methods, for BSPs
of various metallicity. From top to bottom: metallicity $Z=0.03,
0.02, 0.01$ and 0.004.} \label{Fig:hs-w94com-th-ff}
\end{figure*}

\begin{figure*}
\centering{
\includegraphics[bb=77 100 571 690,height=17.1cm,width=17.5cm,clip,angle=-90]{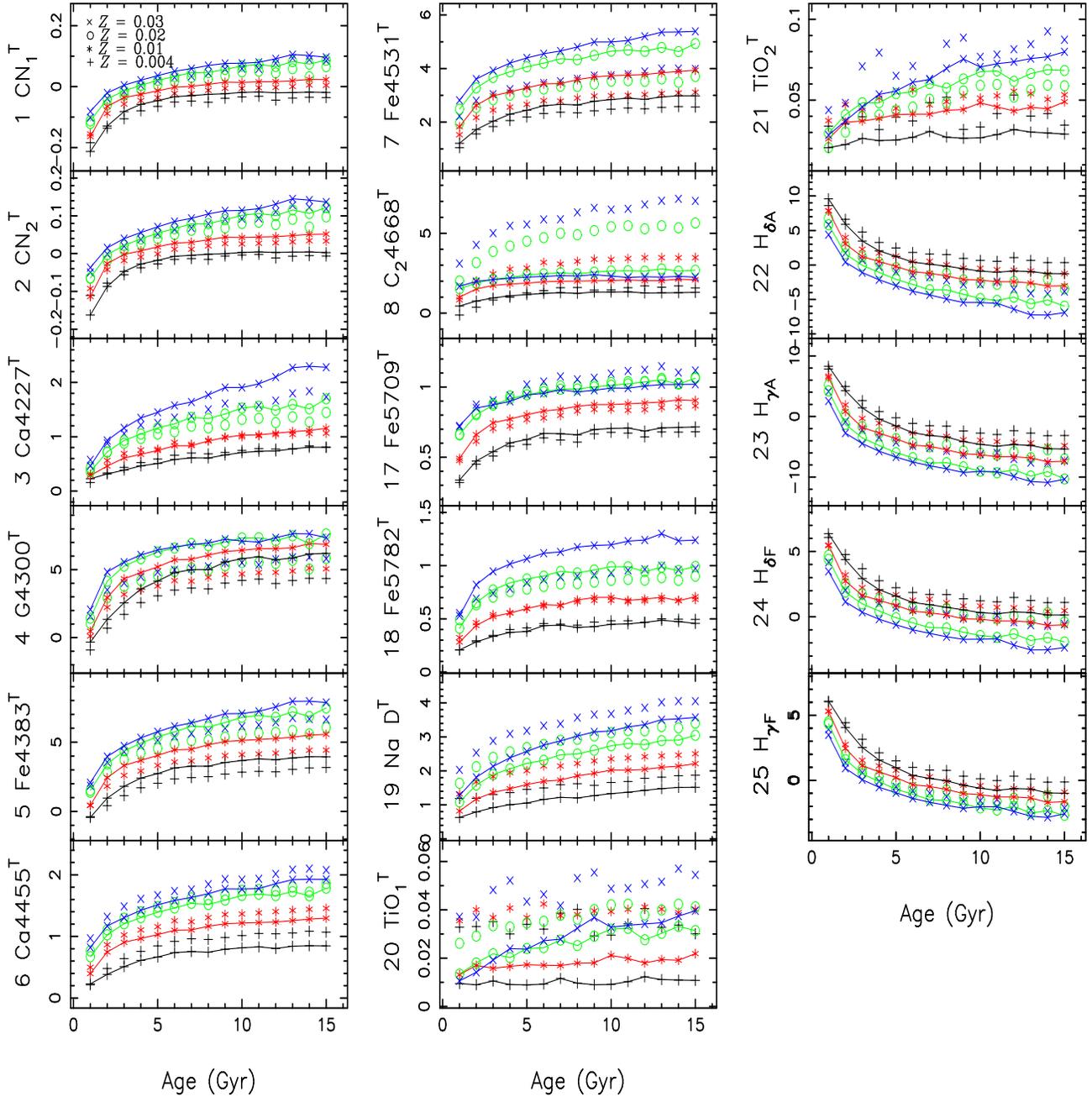}
} \caption{Similar to Fig. \ref{Fig:hs-w94com-th-ff}, but for the
13 refined indices in the T98 system and four Balmer indices of
WO97.} \label{Fig:hs-t98com-th-ff}
\end{figure*}

In Paper II we used two methods to obtain 25 Lick/IDS spectral
absorption-line indices (21 indices of W94 and four Balmer indices
of WO97) for a set of instantaneous-burst BSPs at high resolution:
(1) directly compute them from the synthetic spectrum (hereafter
DC method); (2) obtain them by using the empirical fitting
functions (hereafter FFs method). This part will use the same two
methods to present the 13 refined Lick/IDS indices based on the
T98 system for BSPs. Those indices in common between in the W94
and T98 systems are not given. Combining with Paper II, a Lick/IDS
index database for BSPs at high resolution ($\sim 0.3\, \rm \AA$)
is presented. In addition, this part will compare the synthetic
BSP Lick/IDS indices between using DC and FFs computation methods.

Note that the index names in the T98 system are the same as the
corresponding ones in the W94 system except for index 8, in order
to distinguish them we will give a superscript 'T' to the index
names in the T98 system throughout further analysis. And, we will
follow the older names when the index names of two systems appear
simultaneously.

%2.1
\subsection{Results of the 13 synthetic Lick/IDS Indices of T98 by both DC and FFs methods}
In Table A1 we present index values of BSPs obtained by the FF
method for the 13 Lick/IDS absorption-line indices redefined by
T98. The continuum flux at the mid-point of the 'feature' passband
$f_{i,{\rm C} \lambda}$ (see eqs. 6 and 8 of Paper II) used to
weigh the star have been derived from the high-resolution HRES
stellar spectra of \citet{gon05}. In Table A2 we show the values
of the indices obtained by a direct measurement on the the
high-resolution integrated spectral energy distributions (ISEDs).
The ages of BSPs are at an interval of $1\,$Gyr in the range
1--15\,Gyr and $Z = 0.004, 0.01, 0.02$ and 0.03.

To those who want to mix a BSP with a simple stellar population
(SSP), we also provide the integrated pseudo-continuum ($F_{i,{\rm
C} \lambda} \equiv \Sigma_{k=1}^{n} f_{i,{\rm C} \lambda}$) of
BSPs with various metallicities, which are presented in Table A3.
Tables A1 to A3 are only available online.

%2.2
\subsection{Comparison of the synthetic Lick/IDS indices
between using DC and FFs methods}

To compare the index strengths obtained by DC method with those by
FFs method, we must transform our predictions by DC method to the
Lick/IDS system. We first broaden our high-resolution ISEDs to the
wavelength-dependent Lick/IDS resolution (WO97). Then, we measure
the index strengths in the broaden ISEDs. In Figs.
\ref{Fig:hs-w94com-th-ff} and \ref{Fig:hs-t98com-th-ff} we give
the comparison between the synthetic indices obtained by DC method
and those by FFs method (generated from Table A1) for BSPs of
various metallicity.

Firstly, from Figs. \ref{Fig:hs-w94com-th-ff} and
\ref{Fig:hs-t98com-th-ff} we see that the evolution of TiO$_1$
(index 20) and TiO$_2$ (21) indices in both W94 and T98 systems is
not smooth for two computation methods, and the trend that
increasing the age makes the indices redder is insignificant at
$Z=0.004$ and 0.01. The synthetic Ca4455 (index 6), Fe4668 (8,
except for $\tau = 1\,$ Gyr at $Z=0.01$ and $\tau \le 3\,$Gyr at
$Z=0.004$), Mg$_{\rm b}$ (13), Fe5709 (17, except for $\tau = 2\,$
Gyr at $Z=0.01$ and $\tau =7,8\,$Gyr at $Z=0.004$), NaD (19),
TiO$_1$ (20) and TiO$_2$ (21, except for $Z=0.02$) indices in the
W94 system, the synthetic Ca4455$\rm ^T$ (6$\rm ^T$, except for
$\tau = 1\,$ Gyr at $Z=0.004$ ), C$_2$4668$\rm ^T$ (8$\rm ^T$,
except for $\tau = 1\,$ Gyr at $Z=0.01$ and $\tau \le 4\,$Gyr at
$Z=0.004$), NaD$\rm ^T$ (19$\rm ^T$), TiO$\rm _1^T$ (20$\rm ^T$)
and TiO$_2\rm ^T$ (21$\rm ^T$, except for $Z=0.02$) indices in the
T98 system obtained via DC method are less (bluer) than the
corresponding ones by using FFs method for all metallicities. The
synthetic Ca4227 (index 3), Fe5782 (18), Ca4227$\rm ^T$ (3$\rm
^T$) and Fe5782$\rm ^T$ (18$\rm ^T$) indices are redder at
$Z=0.03$ and become to be bluer at $Z=0.004$, Fe5709$\rm ^T$
(17$\rm^T$) index is bluer at $Z=0.03$ and becomes to be redder at
$Z=0.004$ than the corresponding ones obtained by FFs method.

Comparing  Fig. \ref{Fig:hs-w94com-th-ff} with Fig.
\ref{Fig:hs-t98com-th-ff}, we find that the discrepancies of the
synthetic BSP Lick/IDS indices in the T98 system between two
computation methods are similar to those in the W94 system except
for Fe5709 (17) index. In the T98 system the synthetic Fe5709 (17)
index obtained by FFs method is greater (redder) at $Z = 0.03$
and becomes to be bluer at $Z = 0.004$ (except for $\tau=1\,$Gyr),
while in the W94 system it is greater (redder) for the four
metallicities than that obtained via DC method.

%3__________________________________________________________________
\section{Lick/IDS Indices for BSPs at intermediate resolution ($3\, \rm \AA$)}
In order to compare theoretical results with some spectroscopic
galaxy surveys at intermediate resolution (such as the Sloan
Digital Sky Survey, SDSS), this part will use DC and FFs methods
to present intermediate-resolution ISEDs and Lick/IDS spectral
absorption-line indices for a set of instantaneous-burst BSPs at
$3\, \rm \AA$. The ISEDs are in the wavelength range of
$3000-7000\, \rm \AA$. This set of indices includes 21 indices
based on the W94 system, four Balmer line indices of WO97 and the
13 indices with the new passband definitions of T98. The BSPs are
at an age interval of $1\,$Gyr in the range $1 \le \tau \le 15
\,$Gyr and $Z=0.004, 0.01, 0.02$ and 0.03.

The intermediate-resolution ISEDs are obtained by applying a
Gaussian broadening function to degrade the high-resolution
(0.3\,$\rm \AA$) ISEDs,
\begin{equation}
F_{\lambda,{\rm is}}= {1 \over {\sigma \sqrt{2\pi}}}
\int_{-\infty}^{+\infty} {\rm d}\lambda' F_{\lambda,{\rm hs}} \exp
\Bigl[-{{(\lambda - \lambda')^2} \over {2\sigma^2}}\Bigl]
\end{equation}
where $F_{\lambda,{\rm hs}}$ and $F_{\lambda,{\rm is}}$ are the
high-resolution and the broaden intermediate-resolution synthetic
spectra,
\begin{equation}
\sigma={{\rm FWHM_{is}^2-FWHM_{hs}^2} \over 2.3548}
\end{equation}
Here, ${\rm FWHM_{is}^2}=3\,\rm \AA$ and ${\rm FWHM_{hs}^2} =
0.3\,\rm \AA$ are the FWHM of the intermediate- and
high-resolution spectra.

For BSPs at intermediate resolution, the synthetic Lick/IDS
absorption-line indices are obtained firstly by using the
empirical fitting functions. Instead of directly using the
high-resolution HRES stellar spectral library of \citet{gon05} we
use the broaden stellar spectra ($3\, \rm \AA$) to obtain the
continuum flux at the midpoint of the feature passband, then use
the fitting functions of W94 and WO97 to obtain Lick/IDS
absorption-line indices for BSPs. Similar to the method of
broadening the ISEDs, we obtain the broaden stellar spectra by
using a Gaussian broadening function to degrade the
high-resolution ($0.3 \, \rm \AA$) stellar spectra in the HRES
library. The results of BSP Lick/IDS absorption-line indices
obtained by FFs method are presented in Table B1. Also, we measure
the BSP Lick/IDS absorption-line indices directly from the
synthetic spectra, and give the results in Table B2.

In Table B3 we present the integrated pseudo-continuum for BSPs at
intermediate resolution. Also, Tables B1 to B3 are only available
online.

%4_____________________________________________________________
\section{Summary and Conclusions}
Using the EPS method we present 13 Lick/IDS absorption-line
indices of T98 for an extensive set of instantaneous-burst BSPs at
high resolution ($\sim\,$0.3\,$\rm \AA$), and 38 indices (21
indices of W94, 13 indices of T98 and 4 indices of WO97) at
intermediate resolution ($3\, \rm \AA$). The ages of the
populations are at an interval of $1\,$Gyr in the range 1--15\,Gyr
and $Z= 0.004, 0.01, 0.02, 0.03$. Together with our previous
paper, a library of Lick/IDS spectral absorption-line indices for
BSPs at high and intermediate resolutions is provided. This set of
indices includes 21 indices of W94, four Balmer indices defined by
WO97 and 13 indices with the new passband definitions of T98. All
synthetic indices are obtained by both FFs and DC methods. The
full set of synthetic indices and the integrated pseudo-continuum
can be obtained through http://www.blackwellpublishing.com/... or
from our website (http://www.ast9.com/), or on request from the
first author. Moreover, the full set of the ISEDs can be obtained
from our website.

Moreover, we give the comparison between the synthetic indices
obtained by FFs method and those obtained by DC method for BSPs
with various metallicities, and find that Ca4455 (index 6), Fe4668
(8), Mg$_{\rm b}$ (13), Fe5709 (17), NaD (19), TiO$_1$ (20) and
TiO$_2$ (21, except for $Z=0.02$) in the W94 system, Ca4455$\rm^T$
(6$\rm^T$), C$_2$4668$\rm^T$ (8$\rm^T$), NaD$\rm^T$ (19$\rm^T$),
TiO$_1\rm^T$ (20$\rm^T$) and TiO$_2\rm^T$ (21$\rm^T$, except for
$Z=0.02$) in the T98 system obtained by DC method are less (bluer)
than the corresponding ones obtained by FFs method for all
metallicities. Ca4227 (index 3), Fe5782 (18), Ca4227$\rm^T$
(3$\rm^T$) and Fe5782$\rm^T$ (18$\rm^T$) are greater at $Z=0.03$
and become to be bluer at $Z=0.004$, Fe5709$\rm^T$ (17$\rm^T$)
index is less at $Z=0.03$ and becomes to be redder at $Z=0.004$
than the corresponding ones obtained by FFs method.

%5__________________________________________________________________
\section*{acknowledgements}
This work was funded by the Chinese Natural Science Foundation
(Grant Nos 10303006, 10273020 \& 10433030), by Yunnan Natural
Science Foundation (Grant No. 2005A0035Q) and by the Chinese
Academy of Sciences (KJCX2-SW-T06). We are also grateful to Dr A.
Bressan, the referee, for his useful comments.

{}

\bsp
\label{lastpage}
\end{document}

\appendix
\onecolumn
\section[]{At high resolution ($\sim 0.3\, \rm \AA$), T98
indices (Table A1, A2) and the integrated pseudo-continuum (Table
A3) for instantaneous-burst BSPs with binary interactions.}

\scriptsize
\begin{longtable}{lccccccccccccc}
\caption{At high resolution ($\sim 0.3\, \rm \AA$), T98 indices of
BSPs obtained with the fitting functions (FFs) method.}
\label{Tab:hs-t98-ff} \\
\hline\hline
      &   (1) &  (2)  &   (3)   &   (4)   &    (5)   &   (6)     &   (7)    &   (8)   &   (17)   &    (18)   &   (19)  &     (20)     &    (21)     \\
 Age  &CN$_1^{\rm T}$ &CN$_2^{\rm T}$ &  Ca4227$^{\rm T}$ &  G4300$^{\rm T}$  &  Fe4383$^{\rm T}$  &   Ca4455$^{\rm T}$  &  Fe4531$^{\rm T}$  & C$_2$4668$^{\rm T}$  & Fe5709$^{\rm T}$  &   Fe5782$^{\rm T}$  &  Na D$^{\rm T}$   & $\rm TiO_1^{\rm T}$  & $\rm TiO_2^{\rm T}$ \\
 Gyr  &  Mag  &  Mag  &    \AA  &   \AA   &   \AA    &    \AA    &  \AA     & \AA     &  \AA     &   \AA     &   \AA   & Mag       &  Mag           \\
\hline
\endfirsthead
\caption{continued.}\\
\hline\hline
      &   (1) &  (2)  &   (3)   &   (4)   &    (5)   &   (6)     &   (7)    &   (8)   &   (17)   &    (18)   &   (19)  &     (20)     &    (21)     \\
 Age  &CN$_1^{\rm T}$ &CN$_2^{\rm T}$ &  Ca4227$^{\rm T}$ &  G4300$^{\rm T}$  &  Fe4383$^{\rm T}$  &   Ca4455$^{\rm T}$  &  Fe4531$^{\rm T}$  & C$_2$4668$^{\rm T}$  & Fe5709$^{\rm T}$  &   Fe5782$^{\rm T}$  &  Na D$^{\rm T}$   & $\rm TiO_1^{\rm T}$  & $\rm TiO_2^{\rm T}$ \\
 Gyr  &  Mag  &  Mag  &    \AA  &   \AA   &   \AA    &    \AA    &  \AA     & \AA     &  \AA     &   \AA     &   \AA   & Mag       &  Mag           \\
\hline
\endhead
\hline
\endfoot

\vspace*{0.3pt} \\
\multicolumn{14}{c}{$Z$ = 0.004} \\
 1.  &  -0.184  &  -0.111  &   0.157  &  -0.906  &  -0.370  &   0.214  &   1.044  &  -0.124  &   0.320  &   0.208  &   1.193  &   0.033  &   0.034  \\
 2.  &  -0.136  &  -0.078  &   0.334  &   0.695  &   0.424  &   0.481  &   1.559  &   0.374  &   0.451  &   0.294  &   1.198  &   0.033  &   0.035  \\
 3.  &  -0.105  &  -0.056  &   0.436  &   1.715  &   1.135  &   0.639  &   1.833  &   0.813  &   0.508  &   0.345  &   1.307  &   0.035  &   0.040  \\
 4.  &  -0.079  &  -0.037  &   0.523  &   2.572  &   1.708  &   0.758  &   2.056  &   1.002  &   0.578  &   0.400  &   1.364  &   0.030  &   0.032  \\
 5.  &  -0.065  &  -0.027  &   0.590  &   3.070  &   2.031  &   0.845  &   2.190  &   1.272  &   0.591  &   0.415  &   1.455  &   0.034  &   0.039  \\
 6.  &  -0.051  &  -0.015  &   0.658  &   3.493  &   2.346  &   0.919  &   2.323  &   1.410  &   0.638  &   0.455  &   1.501  &   0.032  &   0.037  \\
\end{longtable}
\begin{longtable}{lccccccccccccc}
\caption{\label{Tab:hs-t98-dir} Similar to Table A1, but for those
 with a direct measurement (DC) method.}\\
\hline\hline
      &   (1) &  (2)  &   (3)   &   (4)   &    (5)   &   (6)     &   (7)    &   (8)   &   (17)   &    (18)   &   (19)  &     (20)     &    (21)     \\
 Age  &CN$_1^{\rm T}$ &CN$_2^{\rm T}$ &  Ca4227$^{\rm T}$ &  G4300$^{\rm T}$  &  Fe4383$^{\rm T}$  &   Ca4455$^{\rm T}$  &  Fe4531$^{\rm T}$  & C$_2$4668$^{\rm T}$  & Fe5709$^{\rm T}$  &   Fe5782$^{\rm T}$  &  Na D$^{\rm T}$   & $\rm TiO_1^{\rm T}$  & $\rm TiO_2^{\rm T}$ \\
 Gyr  &  Mag  &  Mag  &    \AA  &   \AA   &   \AA    &    \AA    &  \AA     & \AA     &  \AA     &   \AA     &   \AA   & Mag       &  Mag           \\
\hline
\endfirsthead
\caption{continued.}\\
\hline\hline
      &   (1) &  (2)  &   (3)   &   (4)   &    (5)   &   (6)     &   (7)    &   (8)   &   (17)   &    (18)   &   (19)  &     (20)     &    (21)    \\
 Age  &CN$_1^{\rm T}$ &CN$_2^{\rm T}$ &  Ca4227$^{\rm T}$ &  G4300$^{\rm T}$  &  Fe4383$^{\rm T}$  &   Ca4455$^{\rm T}$  &  Fe4531$^{\rm T}$  & C$_2$4668$^{\rm T}$  & Fe5709$^{\rm T}$  &   Fe5782$^{\rm T}$  &  Na D$^{\rm T}$   & $\rm TiO_1^{\rm T}$  & $\rm TiO_2^{\rm T}$ \\
 Gyr  &  Mag  &  Mag  &    \AA  &   \AA   &   \AA    &    \AA    &  \AA     & \AA     &  \AA     &   \AA     &   \AA   & Mag       &  Mag           \\
\hline
\endhead
\hline
\endfoot

\vspace*{1pt} \\
\multicolumn{14}{c}{$Z$ = 0.004} \\

 1.  &  -0.213  &  -0.144  &   0.289  &   0.008  &  -0.068  &   0.367  &   1.501  &   0.465  &   0.391  &   0.227  &   0.672  &   0.009  &   0.022  \\
 2.  &  -0.124  &  -0.065  &   0.403  &   1.744  &   1.406  &   0.590  &   2.076  &   0.790  &   0.546  &   0.310  &   0.844  &   0.009  &   0.024  \\
 3.  &  -0.073  &  -0.024  &   0.489  &   2.971  &   2.317  &   0.768  &   2.415  &   0.996  &   0.625  &   0.361  &   0.980  &   0.011  &   0.028  \\
 4.  &  -0.047  &  -0.003  &   0.588  &   3.982  &   2.972  &   0.908  &   2.697  &   1.144  &   0.697  &   0.401  &   1.069  &   0.009  &   0.027  \\
 5.  &  -0.033  &   0.007  &   0.646  &   4.581  &   3.338  &   0.985  &   2.859  &   1.232  &   0.725  &   0.412  &   1.121  &   0.009  &   0.027  \\
 6.  &  -0.019  &   0.018  &   0.742  &   5.196  &   3.799  &   1.079  &   3.027  &   1.321  &   0.775  &   0.465  &   1.238  &   0.009  &   0.029  \\
\end{longtable}
\normalsize
%

%%%%%%%%%%%%%%%%%%%%%%%%%%%%%%%%%%%%%%%%%%%%%%%%%%%%%%%%%%%%%%%%%%%%%%%%%%%%%%%%%%%%
%Table A3
\newpage
\scriptsize
\begin{longtable}{lrrrrrrrrrrrrrrr}
\caption{\label{Tab:is-ifc} At high resolution ($\sim 0.3\,\rm \AA$), the integrated
pseudo-continuum (in units of $10^{34}$ erg.s$^{-1}.{\rm\AA}$) of BSPs.}\\

\hline\hline
Index    &    1Gyr  &    2Gyr  &    3Gyr  &    4Gyr  &    5Gyr  &    6Gyr  &    7Gyr  &    8Gyr  &    9Gyr  &   10Gyr  &   11Gyr  &   12Gyr  &   13Gyr  &    14Gyr &   15Gyr  \\
\hline
\endfirsthead
\caption{continued.}\\
\hline\hline
Index   &    1Gyr  &    2Gyr  &    3Gyr  &    4Gyr  &    5Gyr  &    6Gyr  &    7Gyr  &    8Gyr  &    9Gyr  &   10Gyr  &   11Gyr  &   12Gyr  &   13Gyr  &     14Gyr &   15Gyr  \\
\hline
\endhead
\hline
\endfoot

\vspace*{1pt} \\
\multicolumn{16}{c}{$Z$ = 0.004} \\
$\rm CN_1$      &  13.594&   6.366&   4.320&   3.112&   2.495&   2.044&   1.752&   1.560&   1.277&   1.114&   1.029&   0.974&   0.888&   0.781&   0.733\\
$\rm CN_1^T$    &  13.635&   6.384&   4.332&   3.122&   2.503&   2.052&   1.758&   1.565&   1.282&   1.118&   1.033&   0.978&   0.891&   0.784&   0.736\\
$\rm CN_2$      &  14.251&   6.651&   4.492&   3.227&   2.582&   2.113&   1.809&   1.608&   1.316&   1.148&   1.060&   1.005&   0.915&   0.805&   0.755\\
$\rm CN_2^T$    &  14.525&   6.739&   4.533&   3.250&   2.598&   2.124&   1.818&   1.615&   1.322&   1.152&   1.064&   1.009&   0.919&   0.808&   0.758\\
Ca4227          &  16.353&   7.169&   4.683&   3.317&   2.636&   2.146&   1.832&   1.622&   1.326&   1.154&   1.065&   1.008&   0.920&   0.807&   0.756\\
$\rm Ca4227^T$  &  16.363&   7.168&   4.679&   3.311&   2.631&   2.140&   1.827&   1.618&   1.322&   1.150&   1.062&   1.005&   0.917&   0.805&   0.754\\
\end{longtable}

%%%%%%%%%%%%%%%%%%%%%%%%%%%%%%%%%%%%%%%%%%%%%%%%%%%%%%%%%%%%%%%%%%%%%%%%%%%%%%%%%%%%
\newpage
\section[]{At intermediate resolution ($3\, \rm \AA$), Lick/IDS
indices (Tables B1, B2) and the integrated pseudo-continuum (B3)
for instantaneous-burst BSPs with binary interactions.}

\scriptsize
\begin{longtable}{lrrrrrrrrrrrrrrr}
\caption{\label{Tab:is-all-ff} At intermediate resolution ($3\,
\rm \AA$), Lick/IDS indices of BSPs obtained by the fitting
functions (FFs) method.}\\

\hline\hline
Index    &    1Gyr  &    2Gyr  &    3Gyr  &    4Gyr  &    5Gyr  &    6Gyr  &    7Gyr  &    8Gyr  &    9Gyr  &   10Gyr  &   11Gyr  &   12Gyr  &   13Gyr  &    14Gyr &   15Gyr  \\
\hline
\endfirsthead
\caption{continued.}\\
\hline\hline
Index   &    1Gyr  &    2Gyr  &    3Gyr  &    4Gyr  &    5Gyr  &    6Gyr  &    7Gyr  &    8Gyr  &    9Gyr  &   10Gyr  &   11Gyr  &   12Gyr  &   13Gyr  &     14Gyr &   15Gyr  \\
\hline
\endhead
\hline
\endfoot

\vspace*{1pt} \\
\multicolumn{16}{c}{$Z$ = 0.004} \\
$\rm CN_1$       &  -0.184  &  -0.136  &  -0.106  &  -0.079  &  -0.065  &  -0.051  &  -0.048  &  -0.051  &  -0.038  &  -0.034  &  -0.032  &  -0.044  &  -0.038  &  -0.035  &  -0.036  \\
$\rm CN_1   ^T$  &  -0.184  &  -0.136  &  -0.105  &  -0.079  &  -0.065  &  -0.051  &  -0.048  &  -0.051  &  -0.038  &  -0.034  &  -0.032  &  -0.044  &  -0.038  &  -0.035  &  -0.036  \\
$\rm CN_2$       &  -0.110  &  -0.077  &  -0.055  &  -0.037  &  -0.027  &  -0.015  &  -0.013  &  -0.015  &  -0.006  &  -0.003  &  -0.002  &  -0.011  &  -0.006  &  -0.005  &  -0.006  \\
$\rm CN_2   ^T$  &  -0.111  &  -0.078  &  -0.056  &  -0.037  &  -0.027  &  -0.015  &  -0.013  &  -0.015  &  -0.006  &  -0.004  &  -0.002  &  -0.011  &  -0.007  &  -0.005  &  -0.006  \\
Ca4227           &   0.157  &   0.333  &   0.435  &   0.522  &   0.589  &   0.656  &   0.698  &   0.682  &   0.708  &   0.743  &   0.775  &   0.769  &   0.773  &   0.829  &   0.800  \\
$\rm Ca4227 ^T$  &   0.157  &   0.334  &   0.436  &   0.523  &   0.590  &   0.657  &   0.700  &   0.684  &   0.710  &   0.746  &   0.779  &   0.773  &   0.777  &   0.833  &   0.805  \\
\end{longtable}
\begin{longtable}{lrrrrrrrrrrrrrrr}
\caption{\label{Tab:is-all-dir} Similar to Table B1, but for those
with a direct measurement (DC) method.}\\

\hline\hline
Index    &    1Gyr  &    2Gyr  &    3Gyr  &    4Gyr  &    5Gyr  &    6Gyr  &    7Gyr  &    8Gyr  &    9Gyr  &   10Gyr  &   11Gyr  &   12Gyr  &   13Gyr  &    14Gyr &   15Gyr  \\
\hline
\endfirsthead
\caption{continued.}\\
\hline\hline
Index   &    1Gyr  &    2Gyr  &    3Gyr  &    4Gyr  &    5Gyr  &    6Gyr  &    7Gyr  &    8Gyr  &    9Gyr  &   10Gyr  &   11Gyr  &   12Gyr  &   13Gyr  &     14Gyr &   15Gyr  \\
\hline
\endhead
\hline
\endfoot

\vspace*{1pt} \\
\multicolumn{16}{c}{$Z$ = 0.004} \\
$\rm CN_1$       &  -0.217  &  -0.128  &  -0.078  &  -0.053  &  -0.040  &  -0.027  &  -0.023  &  -0.018  &  -0.016  &  -0.011  &  -0.009  &  -0.012  &  -0.011  &  -0.010  &  -0.011  \\
$\rm CN_1   ^T$  &  -0.214  &  -0.126  &  -0.076  &  -0.050  &  -0.037  &  -0.023  &  -0.019  &  -0.015  &  -0.012  &  -0.007  &  -0.005  &  -0.008  &  -0.007  &  -0.005  &  -0.006  \\
$\rm CN_2$       &  -0.167  &  -0.083  &  -0.037  &  -0.014  &  -0.003  &   0.008  &   0.011  &   0.014  &   0.017  &   0.020  &   0.022  &   0.020  &   0.021  &   0.022  &   0.020  \\
$\rm CN_2   ^T$  &  -0.146  &  -0.067  &  -0.026  &  -0.005  &   0.005  &   0.016  &   0.019  &   0.021  &   0.023  &   0.027  &   0.028  &   0.027  &   0.028  &   0.029  &   0.027  \\
Ca4227           &   0.308  &   0.435  &   0.535  &   0.650  &   0.714  &   0.820  &   0.861  &   0.851  &   0.927  &   0.989  &   1.018  &   1.021  &   1.079  &   1.110  &   1.106  \\
$\rm Ca4227 ^T$  &   0.287  &   0.402  &   0.491  &   0.592  &   0.652  &   0.748  &   0.790  &   0.783  &   0.854  &   0.916  &   0.946  &   0.956  &   1.012  &   1.048  &   1.043  \\
\end{longtable}
\normalsize
%
%
%
%
%
%%%%%%%%%%%%%%%%%%%%%%%%%%%%%%%%%%%%%%%%%%%%%%%%%%%%%%%%%%%%%%%%%%%%%%%%%%%%%%%%%%%%
%Table B3
\scriptsize
\begin{longtable}{lrrrrrrrrrrrrrrr}
\caption{\label{Tab:is-ifc} At intermediate resolution ($3\,\rm
\AA$), the integrated pseudo-continuum of BSPs (in units of
$10^{34}$ erg.s$^{-1}.{\rm\AA}$).}\\

\hline\hline
Index    &    1Gyr  &    2Gyr  &    3Gyr  &    4Gyr  &    5Gyr  &    6Gyr  &    7Gyr  &    8Gyr  &    9Gyr  &   10Gyr  &   11Gyr  &   12Gyr  &   13Gyr  &    14Gyr &   15Gyr  \\
\hline
\endfirsthead
\caption{continued.}\\
\hline\hline
Index   &    1Gyr  &    2Gyr  &    3Gyr  &    4Gyr  &    5Gyr  &    6Gyr  &    7Gyr  &    8Gyr  &    9Gyr  &   10Gyr  &   11Gyr  &   12Gyr  &   13Gyr  &     14Gyr &   15Gyr  \\
\hline
\endhead
\hline
\endfoot

\vspace*{1pt} \\
\multicolumn{16}{c}{$Z$ = 0.004} \\
$\rm CN_1$      &  13.599&   6.367&   4.320&   3.112&   2.494&   2.044&   1.751&   1.559&   1.276&   1.114&   1.029&   0.974&   0.887&   0.781&   0.733\\
$\rm CN_1^T$    &  13.630&   6.378&   4.327&   3.117&   2.498&   2.048&   1.755&   1.562&   1.279&   1.116&   1.031&   0.976&   0.889&   0.783&   0.734\\
$\rm CN_2$      &  14.233&   6.641&   4.485&   3.223&   2.579&   2.110&   1.807&   1.606&   1.315&   1.146&   1.059&   1.004&   0.914&   0.804&   0.754\\
$\rm CN_2^T$    &  14.512&   6.734&   4.530&   3.248&   2.596&   2.122&   1.817&   1.614&   1.321&   1.151&   1.063&   1.008&   0.919&   0.807&   0.757\\
Ca4227          &  16.356&   7.170&   4.684&   3.318&   2.637&   2.146&   1.833&   1.622&   1.326&   1.154&   1.065&   1.008&   0.920&   0.807&   0.756\\
$\rm Ca4227^T$  &  16.382&   7.180&   4.688&   3.320&   2.638&   2.147&   1.833&   1.623&   1.327&   1.154&   1.066&   1.009&   0.920&   0.808&   0.757\\
\end{longtable}
\normalsize